\documentclass[hyper,letterpaper,12pt]{article}
\usepackage{a4wide}
\usepackage{amsmath}
\usepackage[
      colorlinks=true,
      linkcolor=blue,
      urlcolor=blue,
      filecolor=black,
      citecolor=blue,
            ]{hyperref}
\usepackage{color}
\usepackage{graphicx}
\usepackage{amsfonts}
\usepackage{amssymb}
\usepackage{epstopdf}

\def\beq{\begin{equation}}
\def\eeq{\end{equation}}

\begin{document}

\title{\bf \ Holographic Entanglement Entropy in P-wave Superconductor Phase Transition}

\author{\large
~Rong-Gen Cai\footnote{E-mail: cairg@itp.ac.cn}~,
~~Song He\footnote{E-mail: hesong@itp.ac.cn}~,
~~Li Li\footnote{E-mail: liliphy@itp.ac.cn}~,
~~Yun-Long Zhang\footnote{E-mail: zhangyl@itp.ac.cn}\\
\\
\small State Key Laboratory of Theoretical Physics,\\
\small Institute of Theoretical Physics, Chinese Academy of Sciences,\\
\small P.O. Box 2735, Beijing 100190, People's Republic of China\\}
\date{\small April 26, 2012}
\maketitle

\begin{abstract}
\normalsize We investigate the behavior of entanglement entropy
across the holographic p-wave superconductor phase transition in an
 Einstein-Yang-Mills theory with a negative cosmological constant.
 The holographic entanglement entropy is calculated for a strip geometry at
 AdS boundary. It is found that the entanglement entropy undergoes a dramatic change
 as we tune the ratio of the gravitational constant to the Yang-Mills
 coupling, and that the entanglement entropy does behave as the thermal entropy
 of the background black holes. That is, the entanglement entropy
 will show the feature of the second order or first order phase
 transition when the ratio is changed. It indicates that the entanglement entropy is a good probe to
investigate the properties of the holographic phase transition.
\end{abstract}

\section{ Introduction}
As a strong-weak duality, the AdS/CFT
correspondence~\cite{Maldacena:1997re,Gubser:1998bc,Witten:1998qj}
provides a powerful method for studying a strongly interacting
system through its gravity dual which is weakly coupled.
 Especially, it has been used widely to model basic phenomena in condensed matter physics,
  such as superconductivity (superfluidity)~\cite{Herzog:2009xv}, Nernst effect~\cite{Hartnoll:2007ih},
  and non-fermi liquid~\cite{Liu:2009dm}. For more related studies,
  see, for example, Refs.~\cite{Hartnoll:2009sz,McGreevy:2009xe,Sachdev:2012dq} and references therein.

The physical picture behind the holographic superconductor model is
as follows.  As the simplest concrete model, consider an
Einstein-Maxwell-scalar field theory with a negative cosmological
constant. At high enough temperature, the Reissner-Norstr\"om-AdS
(RN-AdS) black hole with a trivial scalar field is stable. And the
dual CFT is in a deconfined phase and describes a conductor phase.
When one lowers the temperature of the black hole, the RN-AdS black
hole becomes unstable, a new black hole solution with nontrivial
scalar field is favored, which can describe a superconducting phase.
The U(1) symmetry is spontaneously broken  due to the nontrivial
scalar field. The condensation  of the scalar ``hair" of the black
hole gives a finite vacuum expectation value of the dual operator in
the field theory side, which plays the role of order parameter in
the holographic phase transition. The s-wave superconductor is
described by the appearance of
  the scalar ``hair"~\cite{Herzog:2009xv,Gubser:2008px,Hartnoll:2008vx,Nishioka:2009zj}, while
  the p-wave superconductor is characterized by the condensation of the
  vector ``hair"~\cite{Gubser:2008zu,Gubser:2008wv}.

On the other hand, the entanglement entropy is expected to be a key
quantity in our understanding some characterization of several
aspects in many-body physics (see, for example,
Refs.~\cite{2006PhRvB..73x5115R,Amico:2007ag}).
 For a given system, the entanglement entropy of subsystem with its complement is defined as the
  von Neumann entropy. In the spirit of AdS/CFT correspondence, a geometric proposal to compute the entanglement
  entropy has been presented in Ref.~\cite{Ryu:2006bv}.
 More precisely, consider a subsystem $\mathcal{A}$ of the total boundary system, the entanglement entropy
 of subsystem $\mathcal{A}$ with its complement is given by looking for the minimal area
  surface $\gamma_\mathcal{A}$ extended into the bulk with the same boundary $\partial\mathcal{A}$
  of $\mathcal{A}$ (see Refs.~\cite{Nishioka:2009un,Takayanagi:2012kg} for reviews)
\begin{equation}\label{law}
S_\mathcal{A}=\frac{\rm Area(\gamma_\mathcal{A})}{4G_N},
\end{equation}
where $G_N$ is the Newton's constant in the bulk. While various
aspects of different holographic superconductor models have been
intensively studied (see, for example,
Refs.~\cite{Albash:2008eh,Hartnoll:2008kx,Cai:2011tm,Horowitz:2008bn,Brynjolfsson:2009ct,Cai:2009hn,
Horowitz:2009ij,Horowitz:2011dz,Montull:2011im,Liu:2012hc,Erdmenger:2011tj}),
the study of entanglement entropy in the holographic phase
transition is just in the early stage. Ref.~\cite{Albash:2012pd}
studied the behavior of entanglement entropy in a holographic s-wave
superconductor model, while Ref.~\cite{Cai:2012sk} discussed the
case in the holographic insulator/superconductor phase transition.

Note that a holographic p-wave superconductor (superfluid) with
fully back reaction in the Einstein-Yang-Mills theory was
constructed in Ref.~\cite{Ammon:2009xh}. This model is interesting
not only because it is a holographic model to describe a p-wave
superconducting phase transition, but also it contains a rich phase
structure.  There is a parameter $\alpha$, the ratio of the
gravitational constant to the Yang-Mills coupling, in this model.
The p-wave superconductor phase transition is second order for small
$\alpha$, while it will become first order as $\alpha$ increases
beyond a critical value.  Therefore it is quite interesting to see
the behavior of entanglement entropy in this model, in particular,
to see how the entanglement entropy changes when the order of the
phase transition changes.

The aim of this paper is just to investigate the behavior of
entanglement entropy in the holographic p-wave superconductor at
finite temperature. The entanglement entropy is calculated for a
straight strip geometry at AdS boundary  by using of the holographic
proposal~\eqref{law}. We find that the behavior of entanglement
entropy changes dramatically when the order of the phase transition
changes. When the strip width is very large, i.e.,
$\gamma_\mathcal{A}$ probes
 deeply, the entanglement entropy is extensive as the thermal entropy of the bulk black hole, while in the opposite
 limit, the behavior perfectly fits the general form obtained from
four-dimensional conformal theories. For the case with an
intermediate strip width, by comparing the entanglement entropy and
the thermal entropy of the bulk black holes during the whole process
of phase transition, we see that they show the same behavior. This
is an interesting and nontrivial result. As a result it shows that
the entanglement entropy is a good probe to the holographic phase
transition and that its behavior can indicate the appearance as well
as the order of phase transition.

The paper is organized as follows. In Section~\eqref{sect:background}, we briefly review the holographic
p-wave superconductor model and give the complete equations of motion to be solved. In
 Section~\eqref{sect:conductor}, the fully back-reacted system is solved by shooting method
 and basic behaviors in equilibrium are described. In Section~\eqref{sect:entropy}, we explore the
 behaviors of the entanglement entropy in the p-wave superconductor phase transition. The conclusion and discussions
  are included in Section~\eqref{sect:conclusion}.


\section{Gravity Background}
\label{sect:background}

We begin with the Einstein-Yang-Mills theory in five-dimensional
asymptotically AdS spacetime
\begin{equation}\label{action}
S =\int d^5 x
\sqrt{-g}[\frac{1}{2\kappa^2}(\mathcal{R}+\frac{12}{L^2})-\frac{1}{4\hat{g}^2}
F^a_{\mu\nu} F^{a\mu \nu}],
\end{equation}
where $\kappa$ is the five dimensional gravitational constant connected with $G_N$ by the relation
$2\kappa^2=16\pi G_N$, $\hat{g}$ is the Yang-Mills coupling constant and $L$ is the AdS radius. The SU(2) Yang-Mills
field strength is
\begin{equation}
 F^a_{\mu\nu}=\partial_\mu A^a_\nu-\partial_\nu A^a_\mu + \epsilon^{abc}A^b_\mu
 A^c_\nu,
\end{equation}
where $\mu,\nu=(t,r,x,y,z)$ denote the indices of spacetime and
$a,b,c=(1,2,3)$ are the indices of the SU(2) group generators
$\tau^a=\sigma^a/2i$ ($\sigma^a$ are Pauli matrices).
$\epsilon^{abc}$ is the totally antisymmetric tensor with
$\epsilon^{123}=+1$. The gauge field is given by
$A=A^a_{\mu}\tau^adx^{\mu}$. Here we define a parameter
$\alpha\equiv\kappa/\hat{g}$ which measures the strength of the back
reaction.

Following Refs.\cite{Gubser:2008wv,Ammon:2009xh}, our ansatz for the metric and Yang-Mills field are chosen by
\begin{equation}\label{metric}
d s^2 = -N(r)\sigma(r)^2 d t^2 + \frac{1}{N(r)} d r^2+r^2 f(r)^{-4} d x^2 + r^2f(r)^2 (d y^2 + d z^2),
\end{equation}
\begin{equation}\label{gauge}
A=\phi(r)\tau^3 dt+ w(r)\tau^1 dx.
\end{equation}
The independent equations of motion in terms of the above ansatz are deduced as follows
\begin{equation}\label{eoms}
 \begin{split}
f''&= -\frac{\alpha^2 f^5 w^2 \phi^2}{3 r^2 N^2 \sigma^2} + \frac{\alpha^2 f^5 {w'}^2}{3 r^2} - f'\left(\frac{3}{r} - \frac{f'}{f} + \frac{N'}{N} +\frac{\sigma'}{\sigma}\right), \\
\phi''&= \frac{f^4 w^2 \phi}{r^2 N} - \phi'(\frac{3}{r} - \frac{\sigma'}{\sigma}), \\
w'' &= -\frac{w \phi^2}{N^2 \sigma^2} - w'\left( \frac{1}{r} + \frac{4 f'}{f} + \frac{N'}{N} + \frac{\sigma'}{\sigma} \right),\\
\sigma'& = \frac{\alpha^2 f^4 w^2 \phi^2}{3 r N^2 \sigma} + \sigma\left(\frac{2 r {f'}^2}{f^2} + \frac{\alpha^2 f^4 {w'}^2}{3 r}\right),\\
m'&=\frac{\alpha^2 r^3 {\phi'}^2}{6\sigma^2} + \frac{r^2 N \sigma'}{2\sigma},
\end{split}
\end{equation}
where $m(r)=\frac{r^4}{2L^2}-\frac{r^2}{2}N(r)$ and `` $'$ " denotes the derivative with respect to $r$. The event horizon $r=r_H$ is determined by the condition $N(r_H)=0$, which gives that $m(r_H)=\frac{r_H^4}{2L^2}$. We should demand $\phi(r_H)=0$ to have a finite form for gauge field at horizon. The asymptotical behavior of these fields near the horizon are
\begin{equation}\label{horizon}
 \begin{split}
 \phi&=\phi_H^{(1)}(1-\frac{r_H}{r})+\phi_H^{(2)}(1-\frac{r_H}{r})^2+\ldots\\\vspace{2mm}
 w&=w_H^{(0)}+w_H^{(1)}(1-\frac{r_H}{r})+\ldots,\\\vspace{2mm}
 {m}&=\frac{r_H^4}{2L^2}+m_H^{(1)}(1-\frac{r_H}{r})+\ldots\\\vspace{2mm}
 {\sigma}&=\sigma_H^{(0)}+\sigma_H^{(1)}(1-\frac{r_H}{r})+\ldots \\\vspace{2mm}
 {f}&=f_H^{(0)}+f_H^{(1)}(1-\frac{r_H}{r})+\ldots.
\end{split}
\end{equation}
All coefficients in above expansions are constants and are related
by the equations of motion~\eqref{eoms}. After substituting the
expansion into~\eqref{eoms}, we find only five independent
parameters, i.e.,
 $\{r_H,\phi_H^{(1)},w_H^{(0)},\sigma_H^{(0)}, f_H^{(0)}\}$.

The ultraviolet (UV) asymptotic expansion near the boundary $r\rightarrow\infty$ behaves as
\begin{equation}\label{boundary}
 \begin{split}
\phi=\phi_B^{(0)}+\frac{\phi_B^{(2)}}{r^2}+\ldots,\quad &w=w_B^{(0)}+\frac{w_B^{(2)}}{r^2}+\ldots,\\\vspace{5mm}
 m=m_B^{(0)}+\frac{m_B^{(2)}}{r^2}+\ldots, \quad \sigma=\sigma_B^{(0)}+&\frac{\sigma_B^{(4)}}{r^4}+\ldots, \quad
 {f}=f_B^{(0)}+\frac{f_B^{(4)}}{r^4}+\ldots.
\end{split}
\end{equation}
To recover the pure AdS boundary, we need the boundary conditions $\sigma_B^{(0)}=1$ and
 $f_B^{(0)}=1$. $\phi_B^{(0)}$ is the chemical potential $\mu$ and $w_B^{(0)}$ is the source
 of the operator $\hat{J}^x_1$. To spontaneously break the U(1) gauge symmetry and rotational
 symmetry, we should impose $w_B^{(0)}=0$.

After imposing boundary conditions, the equations of motion can be solved numerically via
tuning the four independent parameters  $\{\phi_H^{(1)},w_H^{(0)},\sigma_H^{(0)}, f_H^{(0)}\}$
to search for solutions that meet the requirements $\sigma_B^{(0)}=1$, $f_B^{(0)}=1,w_B^{(0)}=0$.
 Notice that the above equations of motion~\eqref{eoms} have four useful scaling symmetries~\cite{Ammon:2009xh}
\begin{equation} \label{scaling1}
\sigma\rightarrow \lambda\sigma,\quad \phi\rightarrow\lambda\phi,
\end{equation}
\begin{equation} \label{scaling2}
f\rightarrow\lambda f,\quad \omega\rightarrow\lambda^{-2}\omega,
\end{equation}
\begin{equation} \label{scaling3}
\quad r\rightarrow \lambda r, \quad m\rightarrow\lambda ^2 m,\quad L \rightarrow\lambda L,
\quad\phi\rightarrow {\lambda^{-1}\phi}, \quad \alpha\rightarrow \lambda \alpha,
\end{equation}
\begin{equation} \label{scaling4}
\quad r\rightarrow \lambda r, \quad \{t,x,y,z\}\rightarrow \lambda^{-1} \{t,x,y,z\},
 \quad m\rightarrow\lambda ^4 m, \quad \omega \rightarrow\lambda \omega,\quad \phi\rightarrow\lambda \phi.
\end{equation}
Taking advantage of the scaling symmetries~\eqref{scaling1} and~\eqref{scaling2}, we will
first choose $\sigma_H^{(0)}=1,f_H^{(0)}=1$ in our shooting method, then use the two scaling
 symmetries again to set $\sigma_B^{(0)}=1,f_B^{(0)}=1$. The last two scaling symmetries allow us to set $L=r_H=1$.



\section{Thermodynamics and Phase Transition}
\label{sect:conductor}
From the discussion in Section \ref{sect:background}, for given
$\{\phi_H^{(1)},w_H^{(0)},\alpha\}$, we can solve the equations of
motion~\eqref{eoms} by choosing $\phi_H^{(1)}$ as a shooting
parameter. After solving the coupled equations, we can obtain the
condensate
$\langle\hat{J}^x_1\rangle=\frac{2\alpha^2}{\kappa^2}w_B^{(2)}$,
chemical potential $\mu$ and total charge density
$\rho=\frac{2\alpha^2}{\kappa^2}\phi_B^{(2)}$ by just reading off
the coefficients $w_B^{(2)}$, $\phi_B^{(0)}$ and $\phi_B^{(2)}$
from~\eqref{boundary} respectively. However, there is an analytic
black hole solution of~\eqref{eoms} for vanishing $\omega(r)$, which
is just the RN-AdS black hole
\begin{equation} \label{RNAdS}
\begin{split}
\phi(r)&=\mu(1-\frac{r_H^2}{r^2}),\quad\omega(r)=0,\quad\sigma(r)=f(r)=1, \\N(r)&=-\frac{2}{r^2}(-\frac{\alpha^2\mu^2 r_H^4}{3}\frac{1}{r^2} +\frac{\alpha^2\mu^2}{3} r_H^2+\frac{1}{2} r_H^4)+r^2.
\end{split}
\end{equation}
This so called RN-AdS solution has vanishing $\omega(r)$, thus
corresponds to the normal phase. From AdS/CFT correspondence, the
boundary thermal equilibrium states are dual to black hole
geometries in bulk, and the Hawking temperature of black hole is
considered as the temperature of the boundary
theory~\cite{Witten:1998zw}. From the metric ansatz~\eqref{metric},
the Hawking temperature of the black hole is
\begin{equation} \label{temp}
T=\frac{\sigma N'}{4 \pi}\Big|_{r=r_H}=
\Big(\frac{\sigma}{\pi L^2}-\alpha^2\frac{\phi'^2}{12\pi \sigma}\Big)r \Big|_{r=r_H}.
\end{equation}
The Bekenstein-Hawking entropy of this black hole is
\begin{equation}\label{BHentropy}
S_T=\frac{A_H}{4G_N}=\frac{2\pi}{\kappa^2}V r_H^3,
\end{equation}
where $A_H$ denotes the area of the horizon and $V=\int dxdydz$.

We will work in the grand canonical ensemble where the chemical
potential $\mu$ at the boundary is fixed. It is convenient to
express physical quantities in scale invariant way. Under the
scaling symmetry~\eqref{scaling4}, the relevant quantities scale as
follows
\begin{equation}
\mu\rightarrow\lambda\mu,\ \ T\rightarrow\lambda T,\ \ \rho\rightarrow\lambda^3\rho,
\ \ \langle\hat{J}^x_1\rangle\rightarrow\lambda^3\langle\hat{J}^x_1\rangle.
\end{equation}
Therefore, we choose the following scale invariant combinations to examine physics in
the grand canonical ensemble
\begin{equation}
\frac{T}{\mu},\ \ \frac{\rho}{\mu^3},\ \ \frac{\langle\hat{J}^x_1\rangle}{\mu^3}.
\end{equation}
\begin{figure}[h]
\centering
\includegraphics[scale=0.80]{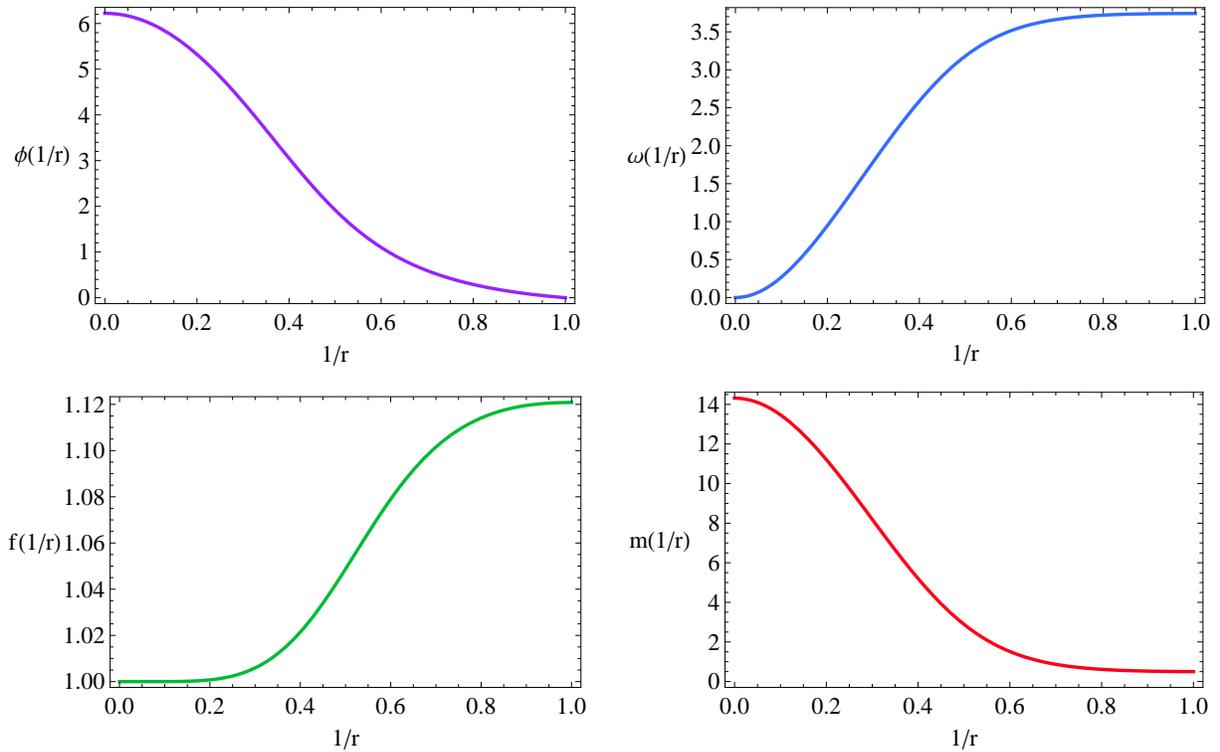}
\caption{\label{functions} The configurations of gauge fields $(\phi,\omega)$
and metric functions $(f,m)$ as a function of the inverse holographic coordinate $1/r$ for
 $\alpha=0.447$ at $T/\mu\simeq0.0215$.}
\end{figure}
In gauge/gravity duality the grand potential $\Omega$ of the
boundary thermal state is identified with $T$ times the on-shell bulk
action in Euclidean signature. The Euclidean action must include the
Gibbons-Hawking boundary term for a well-defined Dirichlet
variational principle and further a surface counterterm for removing
divergence
\begin{equation}
S_{Euclidean}=-\int d^5 x
\sqrt{g}[\frac{1}{2\kappa^2}(R+\frac{12}{L^2})-\frac{1}{4\hat{g}^2}
F^a_{\mu\nu}
F^{a\mu\nu}]+\frac{1}{2\kappa^2}\int_{r\rightarrow\infty} d^4x
\sqrt{h}(-2K+\frac{6}{L^2}),
\end{equation}
where $h$ is the induced metric on the boundary $r\rightarrow\infty$, and $K$ is the trace of the extrinsic
curvature.

This model has been numerically solved in Ref.~\cite{Ammon:2009xh} and shown that the order of the phase transition
relies on the value of $\alpha$. The transition is second order as $\alpha$ is less than $\alpha_c=0.365\pm0.001$,
while it is first order for larger values than $\alpha_c$.
We will re-solve the equations of motion~\eqref{eoms} for completeness and for further discussion. Typical
solutions for the metric and gauge field configurations are presented in~Figure.\eqref{functions}, which are
 needed to calculate entanglement entropy in the next section.

To compare the differences between the second order transition and first order transition,
we choose $\alpha=0.316<\alpha_c$ and $\alpha=0.447>\alpha_c$ as concrete examples.
\begin{figure}[h]
\centering
\includegraphics[scale=0.93]{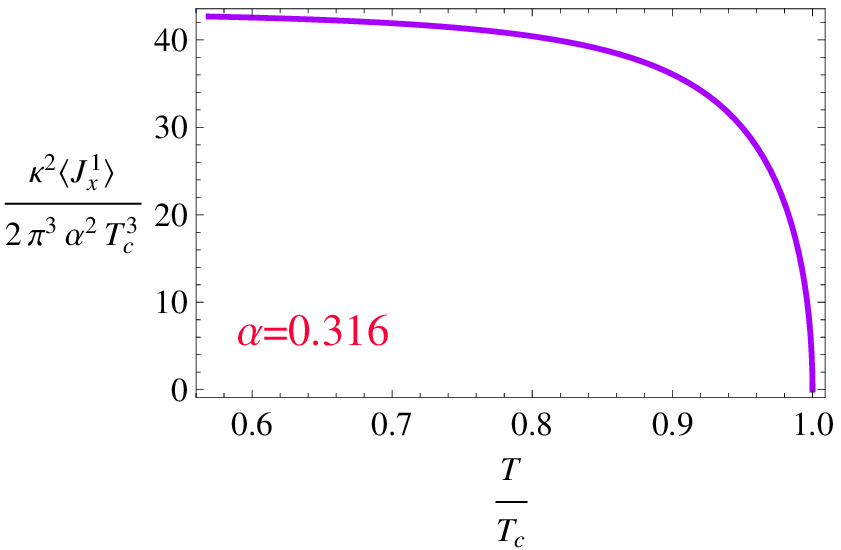}
\includegraphics[scale=0.75]{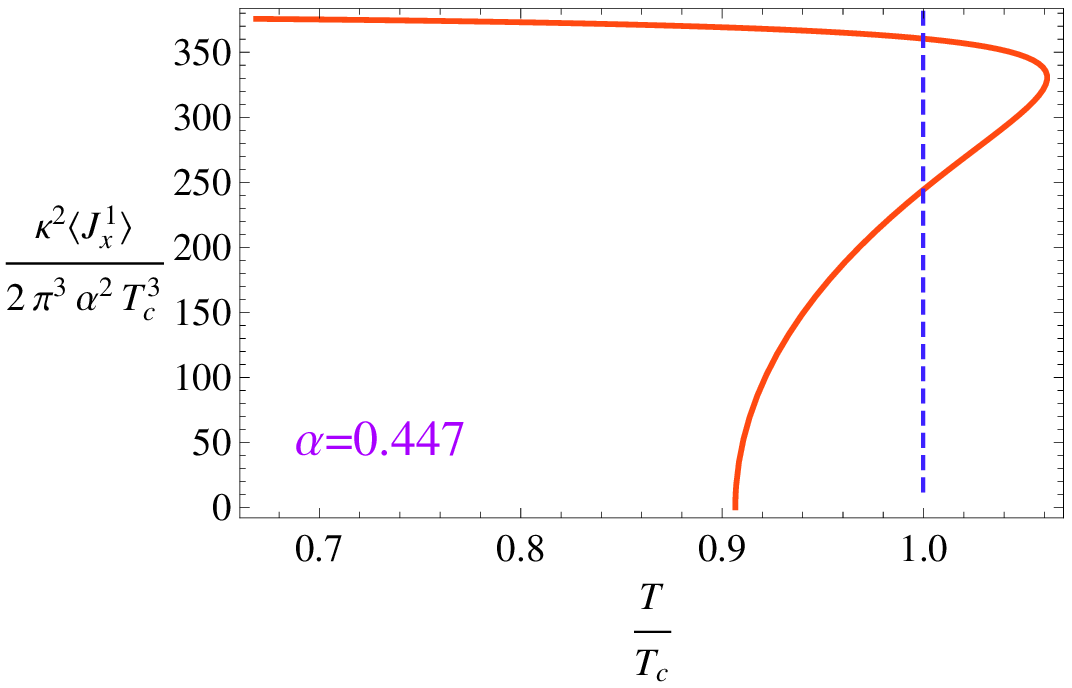}
\caption{\label{condensate} The condensation of vector operator $\hat{J}^x_1$
as a function of temperature for $\alpha=0.316$ (left plot) and $\alpha=0.447$ (right plot). The vertical dashed line on the right plot represents the transition temperature. The right plot has a region where $\langle\hat{J}^x_1\rangle$ is multi-valued.}
\end{figure}
The condensation of vector operator $\hat{J}^x_1$ as a function of temperature for $\alpha=0.316$ and
 $\alpha=0.447$ is displayed in~Figure.\eqref{condensate}. For the case $\alpha=0.316$,
 the condensate $\langle\hat{J}^x_1\rangle$ appears at a particular temperature $T_c\simeq0.0458\mu$. As the
 temperature is lowered, $\langle\hat{J}^x_1\rangle$ increases continuously. The critical behavior near $T_c$
 is found to be $\langle\hat{J}^x_1\rangle\propto(1-T/T_c)^{\frac 1 2}$, which is the typical result from the
 mean-field theory. While for the case $\alpha=0.447$, we can see from~Figure.\eqref{condensate} that
 the curve of the condensate has two branches when $0.905<T/T_c<1.06$. Therefore, the value of condensate
  has a jump at critical temperature $T_c\simeq0.0218\mu$, which represents a first order phase transition.

To distinguish which branch is physical, i.e., thermodynamically
favored, we need to calculate the
 grand potential $\Omega$.
\begin{figure}[h]
\centering
\includegraphics[scale=0.92]{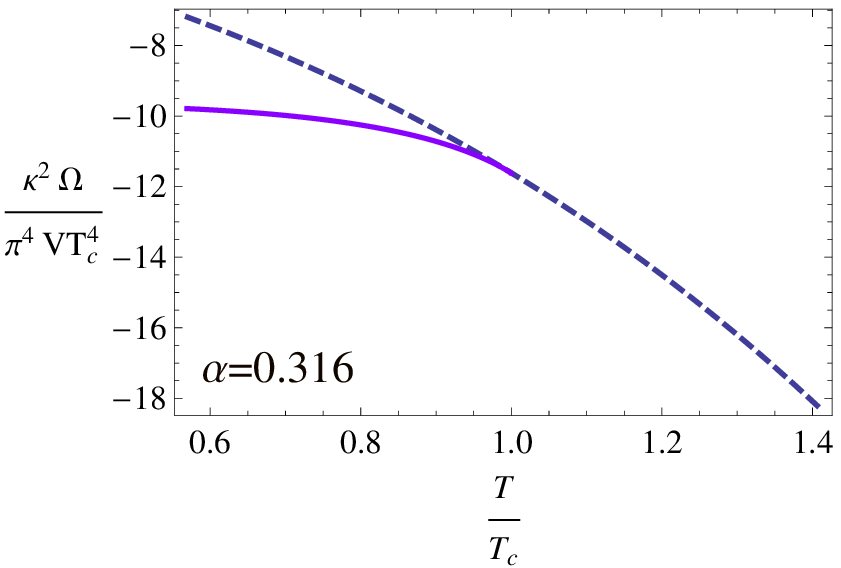}
\includegraphics[scale=0.94]{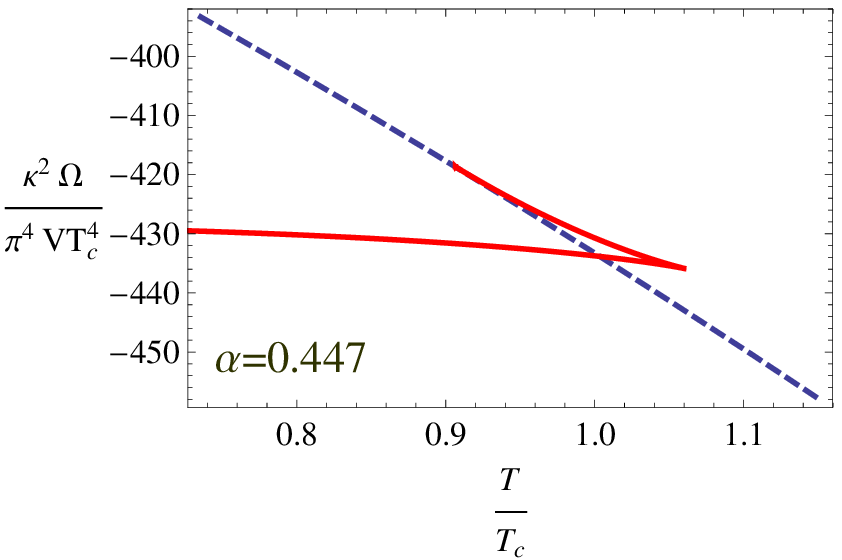}
\caption{\label{potential} The grand potential $\Omega$ as a
function of temperature for $\alpha=0.316$ (left plot) and
$\alpha=0.447$ (right plot). Trace the physical curve by choosing
the lowest grand potential at a fixed $T$. The critical temperature
$T_c$ is the point at which the superconductor phase begins to be
thermodynamically preferred. In both plots, dashed blue curves are
for the RN-AdS solutions, while the solid curves are for the
superconductor solutions.}
\end{figure}
\begin{figure}[h]
\centering
\includegraphics[scale=1.10]{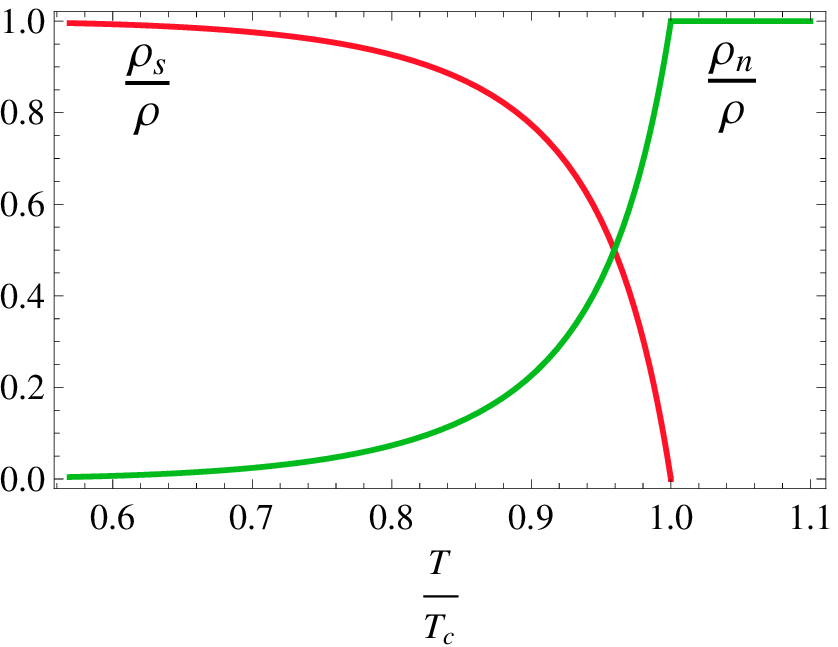}
\caption{\label{ratio} $\frac{\rho_s}{\rho}$ and
$\frac{\rho_n}{\rho}$ are plotted versus  temperature in
$\alpha=0.316$ case. In our numerical calculation, we find
$\frac{\rho_s}{\rho}$ goes to zero linearly near $T_c$, which is the
same as the critical behavior obtained in the probe limit.}
\end{figure}
The values of grand potential $\Omega$ are exhibited
in~Figure.\eqref{potential}. The RN-AdS solutions always exist for
all temperatures, but it is only thermodynamically favored at high
temperatures $T>T_c$. As the temperature is lowered below critical
value for each $\alpha$, the grand potential $\Omega$ from
superconductor solution is smaller than the one from the RN-AdS
solution, thus the superconductor phase is physically compared to
the normal phase (RN-AdS solution). Therefore, there is a phase
transition occurring at $T_c$. For $\alpha=0.316$ case, the
transition is second order. However, for $\alpha=0.447$ case, there
is a characteristic ``swallowtail" shape of the grand potential,
signaling  a first order transition.

According to the two-fluid model, the total charge density $\rho$ can be divided into
two components $\rho=\rho_n+\rho_s$, where $\rho_n$ is the normal component, while $\rho_s$
is the superconducting component. In the holographic setup, the normal charge density $\rho_n$ is
proportional to the $\tau^3$ part of the electric field at the horizon~\cite{Gubser:2008wv}, which
is given by $\frac{\alpha^2}{\kappa^2} \phi_H^{(1)}$ in our units. Therefore the superconducting charge
density is $\rho_s=\rho-\rho_n$, where $\rho=\frac{2\alpha^2}{\kappa^2}\phi_B^{(2)}$ is the total charge density.
 We draw $\frac{\rho_s}{\rho}$ and $\frac{\rho_n}{\rho}$ versus temperature in
 Figure.\eqref{ratio}. $\frac{\rho_s}{\rho}$ and $\frac{\rho_n}{\rho}$ as two functions of temperature
  are reminiscent of the temperature dependence of the superfluid and normal components of liquid
  He~$\mathrm{II}$ as measured from in the torsional oscillation disk stack experiment. $\rho_s$ seems to
  vanish near $T_c$ like a power law, i.e., $(T_c-T)^\nu$. However, we find $\frac{\rho_s}{\rho}$ goes to
  zero linearly here, while the experiment gives a value of $\nu\simeq0.67$.


\section{Entanglement Entropy}
\label{sect:entropy}

After solving the equations of motion, we are now ready to calculate the entanglement entropy in this
holographic model.
 Because of the arbitrary choice of the subsystem $\mathcal{A}$, we can define
infinite entanglement entropies correspondingly. However, we are
here interested in a belt geometry with a finite width $\ell$ along
the $x$ direction and infinitely extending in $y$ and $z$
directions.

\begin{figure}[h]
\centering
\includegraphics[scale=1.20]{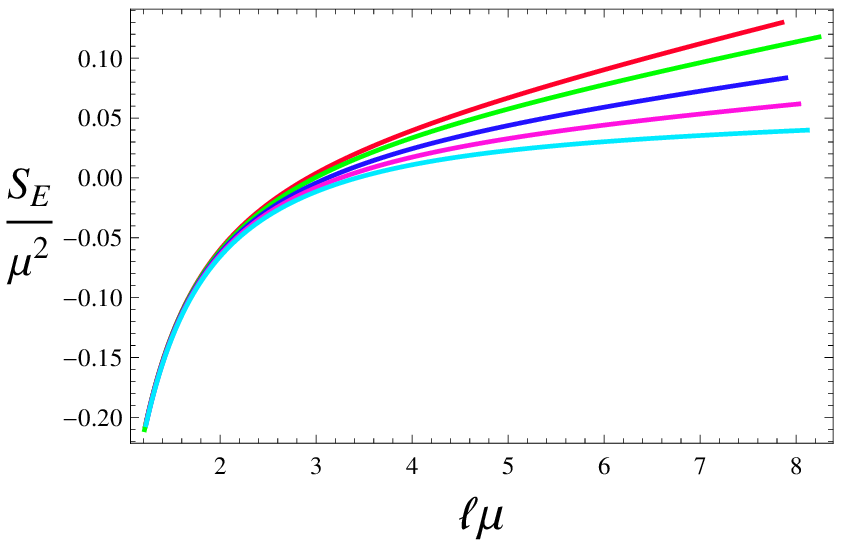}
\caption{\label{Entropywidth2} The entanglement entropy in
$\alpha=0.316$ case, as a function of belt width at fixed
temperature. The curves from top to bottom correspond to
$T\simeq0.0458\mu,0.0448\mu,0.0424\mu, 0.0387\mu$, and $0.0294\mu$,
respectively.}
\end{figure}
To deal with the UV divergence, we assume that the subsystem
$\mathcal{A}$ sites on the slice $r=\frac{1}{\epsilon}$ where
$\epsilon\rightarrow0$ is the UV cutoff. More specifically,
$\gamma_A$ starts from $x=-\frac{\ell}{2}$ at
$r=\frac{1}{\epsilon}$, extends into the bulk until it reaches the
minimum $r=r_*$, then returns back to the AdS boundary
$r=\frac{1}{\epsilon}$ at $x=+\frac{\ell}{2}$. According to the
proposal~\eqref{law}, we need to minimize the following area
functional
\begin{equation}\label{surface}
Area(\gamma_\mathcal{A})={V_2}\int_{-\frac{\ell}{2}}^{+\frac{\ell}{2}}dx\sqrt{\frac{r^4f(r)^4}{N(r)}(\frac{dr}{dx})^2+r^6},
\end{equation}
where $V_2=\int dydz$. The integrand can be considered as the
Lagrangian with $x$ direction thought of as time. As the Lagrangian
does not explicitly depend on ``time $x$", the Hamiltonian is
conserved. Thus we can easily deduce the ``equation of motion" that
gives minimal area from \eqref{surface}
\begin{equation}\label{minimal}
\frac{dr}{dx}=\pm \frac{r \sqrt{N(r)}}{f(r)^2}\sqrt{\frac{r^6}{r_*^6}-1},
\end{equation}
where we demand that the surface is smooth at the turning point
$r=r_*$, i.e., $dr/dx|_{r=r_*}=0$. Integrated once, the belt width
$\ell$ can be fixed as
\begin{equation}\label{width}
\frac{\ell}{2}=\int_{r_*}^{\frac{1}{\epsilon}}dr\frac{dx}{dr}=\int_{r_*}^{\frac{1}{\epsilon}}
dr\frac{f(r)^2}{r\sqrt{N(r)}}\frac{r_*^3}{\sqrt{r^6-r_*^6}}.
\end{equation}
Substituting~\eqref{minimal} into~\eqref{surface}, we finally obtain the entanglement entropy
\begin{equation}\label{entropy}
S_E=\frac{V_2}{2G_N}\int_{r_*}^{\frac{1}{\epsilon}}dr\frac{r^5f(r)^2}{\sqrt{N(r)}}\frac{1}{\sqrt{r^6-r_*^6}}
=\frac{2\pi}{\kappa^2}V_2(\frac{1}{\epsilon^2}+S_E),
\end{equation}
where the UV cutoff $1/\epsilon$ has been taken into consideration.
The first term indicates UV divergent
 ($\epsilon\rightarrow 0$) and represents the ``area law"~\cite{Ryu:2006bv,Srednicki:1993im}.
 It can be deduced by plugging the UV asymptotic expansion~\eqref{boundary} into~\eqref{entropy}.
  While the second term is independent of the cutoff and is finite, so this term is physical important.
  Following the discussion in Section~\eqref{sect:conductor}, $\ell$ and $S_E$ under the
  transformation~\eqref{scaling4} scale as $\ell\rightarrow\lambda^{-1}\ell,\ \ S_E\rightarrow\lambda^2 S_E$,
  so we will introduce the scale invariants
\begin{equation}
\mu\ell,\ \ \frac{S_E}{\mu^2}.
\end{equation}

\begin{figure}[h]
\centering
\includegraphics[scale=0.94]{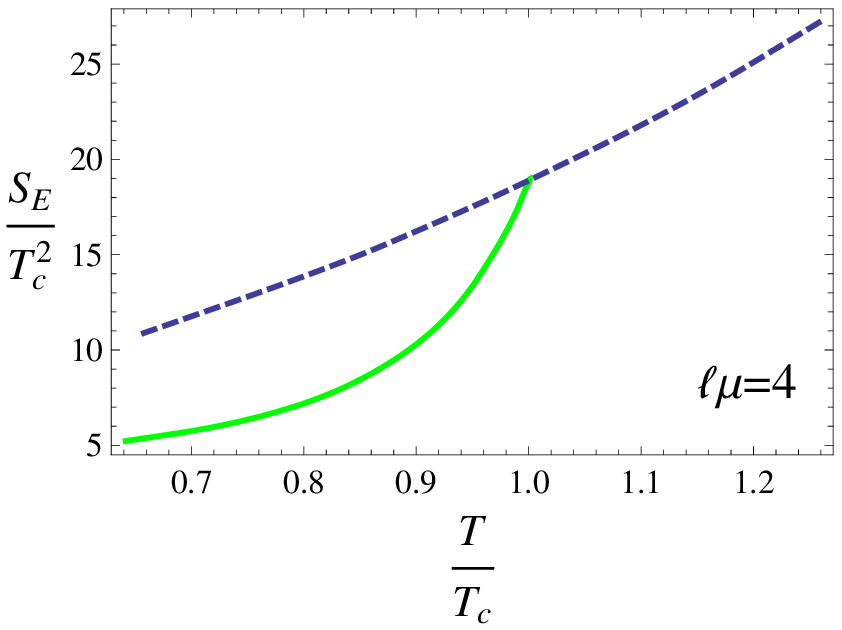}
\includegraphics[scale=0.94]{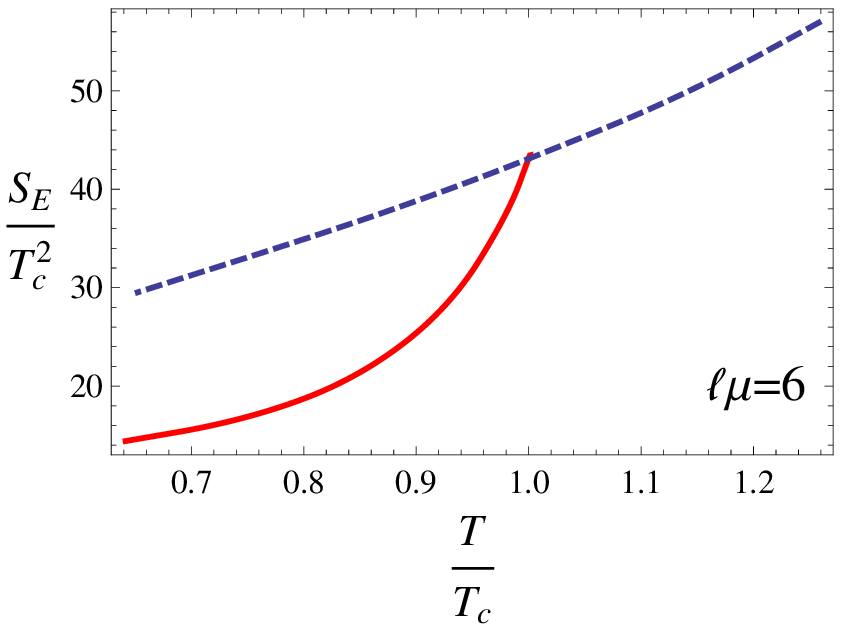}
\caption{\label{Entropy2} The entanglement entropy in $\alpha=0.316$
case, as a function of temperature at fixed belt width (left plot
for $\ell\mu=4$ and right plot for $\ell\mu=6$). The dashed blue
curves are from the RN-AdS solutions, while the solid curves are
from superconductor solutions. The physical curve is determined by
selecting the lower entropy at a given $T$.}
\end{figure}

We first focus on the case with second order phase transition, i.e.,
$\alpha<\alpha_c$. In~Figure\eqref{Entropywidth2} we plot the
behavior of the universal part of the entanglement entropy $S_E$ as
a function of belt width $\ell$ by fixing the temperature. The curve
at the top is at the transition temperature $T_c$, which is
identical with the RN-AdS case.  We observe that the slope of the
curve decreases as  the temperature is lowered in superconductor
situation as is expected that the lower the temperature is, the more
the degrees of freedom will condense. This phenomenon can be seen
much more clearly in~Figure\eqref{Entropy2}, which shows how the
entanglement entropy evolves with temperature by fixing the belt
width. We see that although the entanglement entropy is continuous
at critical temperature $T_c$, there is a discontinuous change in
its slope at $T_c$. This discontinuity may signal a significant
reorganization of the degrees of freedom of the system, since some
kind of new degrees of freedom, like the Cooper pair, would emerge
in the new phase.

\begin{figure}[h]
\centering
\includegraphics[scale=0.87]{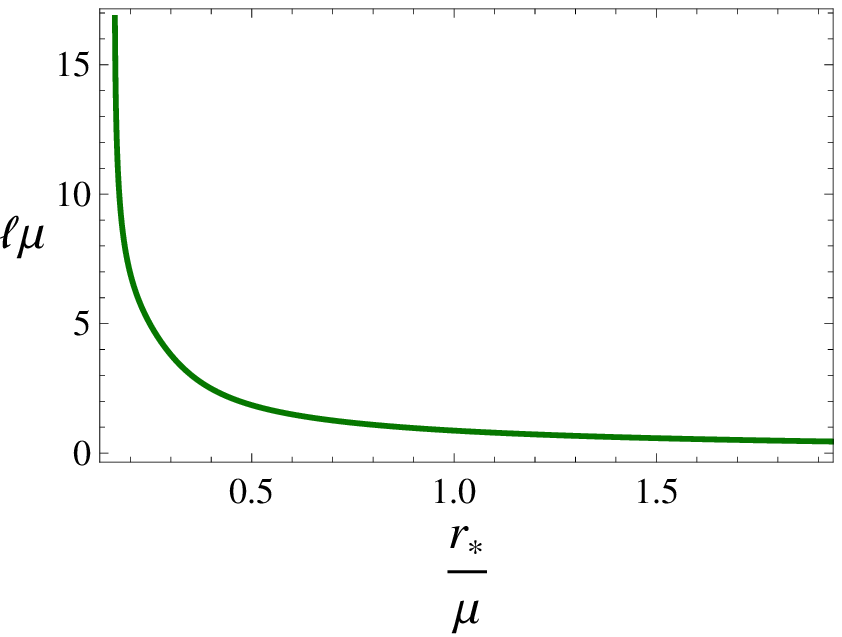}
\includegraphics[scale=1.0]{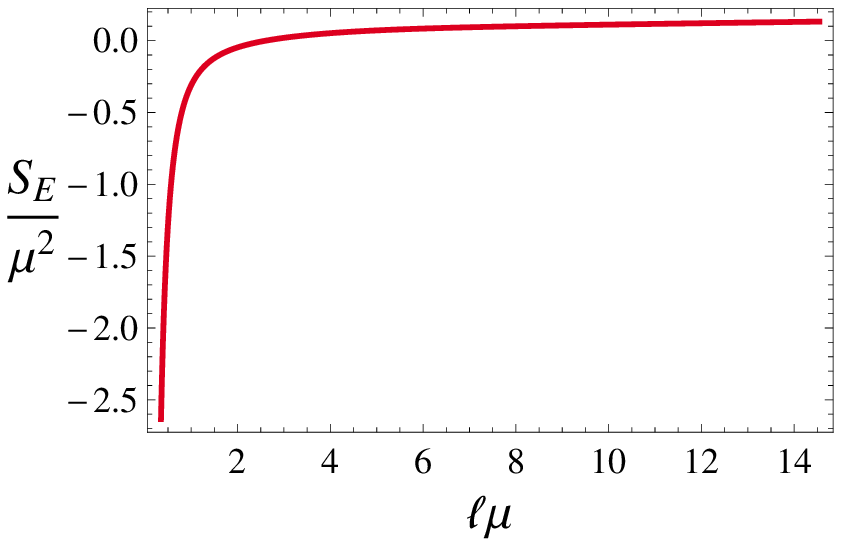}
\caption{\label{Entropywidth1} The left plot shows the behavior of
belt width as a function of turning point $r_*$ for $\alpha=0.447$
at $T/\mu\simeq0.0215$. While the entanglement entropy as a function
of belt width is presented in the right plot. }
\end{figure}

We show the results for the first order transition case in~Figure\eqref{Entropywidth1}. The behavior of entanglement
entropy at fixed temperature is quite similar to the second order transition case. $S_E$ changes monotonously
 with respect to the belt width. This behavior is quite different from the result in Ref.~\cite{Albash:2012pd},
  where a swallowtail shape appears in a region with finite belt
  width. It can be seen from the left plot in
  Figure\eqref{Entropywidth1} that the belt width monotonously
  decreases as the turning point $r_*$ increases. Actually, $\ell$ diverges logarithmically as $r_* \rightarrow r_H$ at
  nonvanishing temperature as observed from~\eqref{width}. In fact in Ref.~\cite{Albash:2012pd}, the non-monotonousness
  of $\ell$ with respect to the turning point $r_*$ is
  indispensable for the emergence of the kink. On this we will have
  more discussions in the last section of the paper.
\begin{figure}[h]
\centering
\includegraphics[scale=0.94]{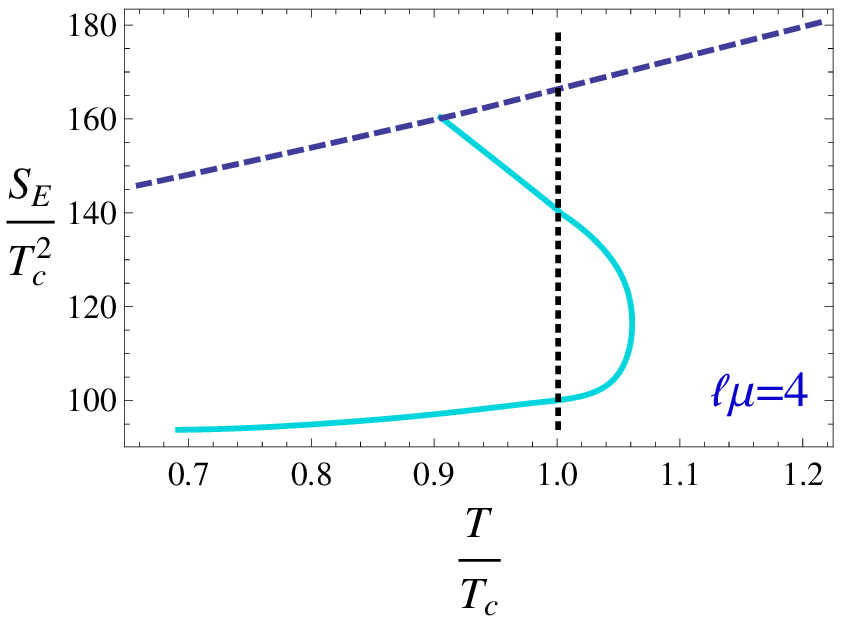}
\includegraphics[scale=0.94]{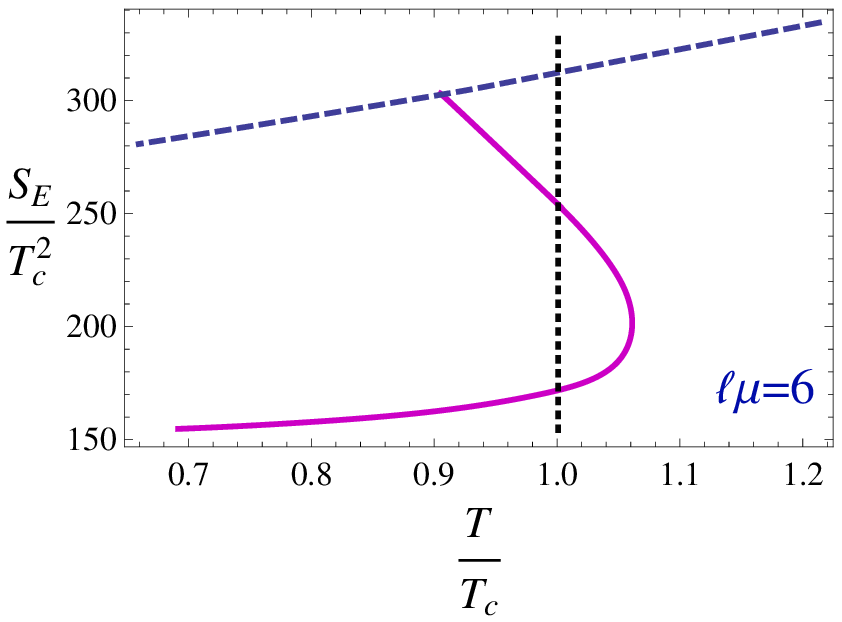}
\caption{\label{Entropy1} The entanglement entropy in $\alpha=0.447$ case,  as a function of temperature at
fixed belt width (left plot for $\ell\mu=4$ and right plot for $\ell\mu=6$). The dashed blue curves are
 from the RN-AdS solutions, while the solid curves are from the superconductor solutions. The physical curve
 is determined by choosing the dashed blue curve above $T_c$, indicated by the vertical dotted line,
  and the curve which has the lowest entropy below $T_c$.}
\end{figure}
\begin{figure}[h]
\centering
\includegraphics[scale=0.945]{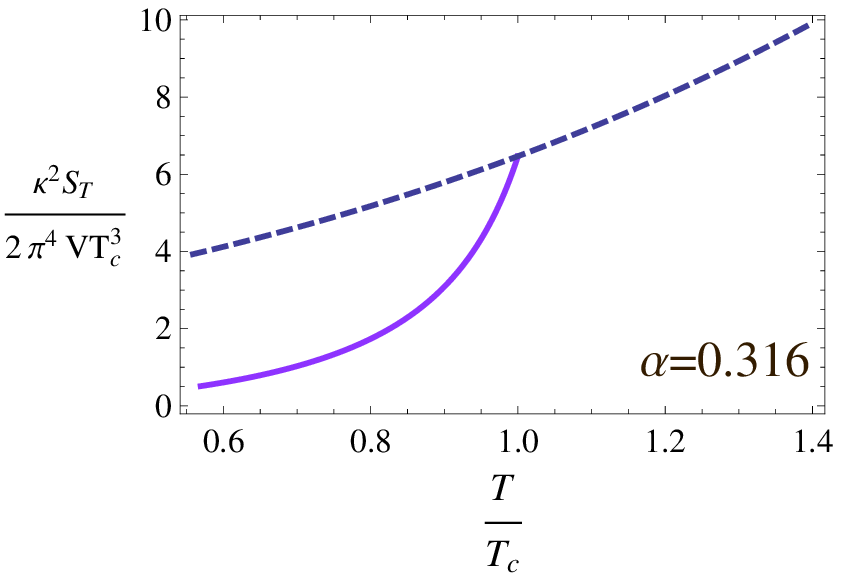}
\includegraphics[scale=0.945]{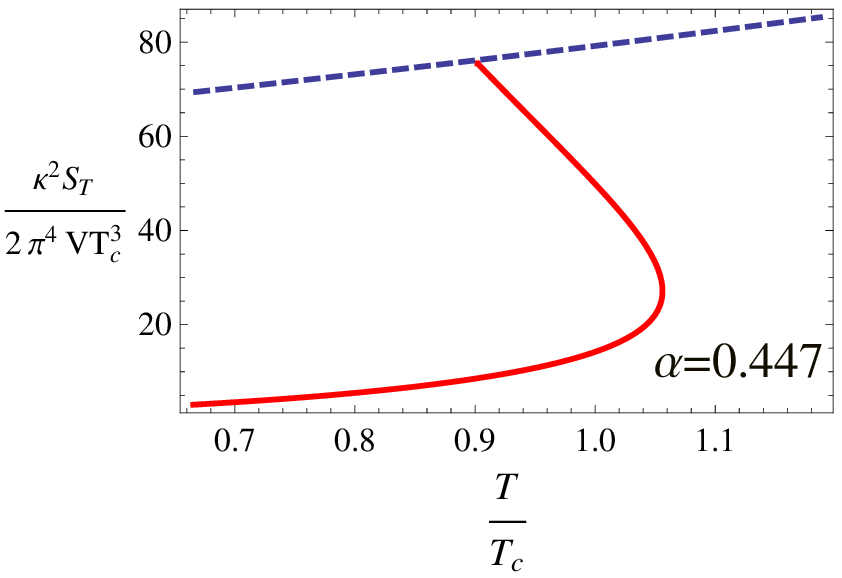}
\caption{\label{TEntropy} The thermal entropy $S_T$ as a function of
temperature for $\alpha=0.316$ (left plot) and $\alpha=0.447$ (right
plot). The dashed blue curves are from the RN-AdS solutions, while
the solid curves are from the superconductor solutions. A
discontinuous slope at $T_c$ in the left plot obviously indicates a
second-order transition, while the jump in the right plot indicates
a first-order transition.}
\end{figure}

The behavior of $S_E$ as a function of temperature for fixed belt
width is presented in~Figure\eqref{Entropy1}. Comparing
with~Figure\eqref{Entropy2}, we can find a dramatic change in the
$\alpha > \alpha_c$ case. The blue curve from normal phase is
physical as $T>T_c$,
 while the curve with the lowest entropy at a given temperature is preferred below $T_c$.
 Therefore, there is an obvious jump in $S_E$ as well as its slope at critical temperature.
  It seems reasonable to expect an abrupt reduction in the number of degrees of freedom at
   $T_c$ since the condensate has a sharp jump at the critical point.

In both two kinds of phase transition, we observe from~Figure.\eqref{Entropywidth2} and~Figure.\eqref{Entropywidth1}
 that $S_E$ exhibits linear behavior with respect to $\ell$ for large $\ell$. Indeed, we can find that in
 large $\ell\mu\sim\ell T$ limit, the main contribution of the integrals~\eqref{width} and~\eqref{entropy} to
  $S_E$ comes from the region near $r=r_*\sim r_H$. In addition to $N(r)\propto(r-r_H)$ near horizon at
   nonvanishing temperature, we can deduce the linear relation $\frac{S_E}{\mu^2} \sim \ell\mu$. Thus we
    obtain the entanglement entropy in this limit
\begin{equation}\label{entropylimit1}
S_E\sim\frac{2\pi}{\kappa^2}\mu^3
V_2\ell\sim\frac{2\pi}{\kappa^2}T^3V,
\end{equation}
where we have subtracted the UV divergent term. This equation is similar to the Bekenstein-Hawking
entropy~\eqref{BHentropy} and may seem to be surprised at first glance since the
entropy~\eqref{entropylimit1} is proportional to the area of the subsystem $\mathcal{A}$ as opposed to
the area law~\eqref{law}. A geometric interpretation made in Ref.~\cite{Ryu:2006bv} is that
 $\gamma_\mathcal{A}$ will wrap a part of the black hole horizon as the increase of belt
 width and therefore is equal to the fraction of black hole entropy .

In contrast, from~Figure.\eqref{Entropywidth2},
and~Figure.\eqref{Entropywidth1}, the value of $\frac{S_E}{\mu^2}$
seems to be power-law divergent as $\mu\ell$ vanishes. In fact, the
relationship between entanglement entropy and belt width in small
$\ell\mu\sim\ell T$ limit is found  to perfectly agree with a
universal function in our numerical calculation
\begin{equation}\label{entropylimit2}
S_E=\frac{2\pi}{\kappa^2}
V_2(\frac{1}{\epsilon^2}-\frac{0.32066}{\ell^2})
=\frac{2\pi}{\kappa^2}T^2V_2(\frac{1}{\epsilon^2
T^2}-\frac{0.32066}{\ell^2T^2}).
\end{equation}
Notice that entanglement entropy in four-dimensional conformal
theories for a belt configuration takes the universal
form~\cite{Casini:2005zv,Nishioka:2006gr}
\begin{equation}\label{UniSentropy}
S_\mathcal{A}=\zeta\frac{V_2}{\epsilon^2} -C\frac{V_2}{2\ell^2},
\end{equation}
where $\zeta$ and $C$ are numerical constants which depend on the
details of a theory under consideration. Our $\ell T\rightarrow0$
result~\eqref{entropylimit2} has the same form
as~\eqref{UniSentropy},
 which is reasonable since our metric solution is asymptotically AdS and $\gamma_\mathcal{A}$ with small
  belt length can only probe the bulk sufficiently near the boundary $r\rightarrow\infty$. As
  discussed in~\cite{Swingle:2011mk}, there exist some universal crossover functions connecting the
   universal parts of the entanglement entropy to the thermal entropy. We try to construct
   the crossover function here as $S_E(\ell,T)=T^n F(\ell T)$. In our situation,
   as $\ell T\rightarrow\infty$, we obtain that n=2 and $F(\xi\rightarrow\infty)\sim\xi$,
   thus $S_E(\ell,T)$ behaves as the extensive thermal entropy. While as $\ell T\rightarrow0$,
    we find $n=2$ and $F(\xi\rightarrow0)\sim\frac{1}{\xi^2}$. For the intermediate scale,
    the precise form of $F(\xi)$ can only be obtained numerically.

To get further understanding of the connection between the
entanglement entropy and thermal entropy, it would be irradiative
and instructive to compare the behaviors of them during the process
of phase transition. Figure.\eqref{TEntropy} shows the thermal
entropy $S_T$ as a function of temperature. The quite qualitatively
similarity in~Figure.\eqref{Entropy2},~Figure.\eqref{Entropy1}
and~Figure.\eqref{TEntropy} is impressive and striking. Note that
$\ell\mu=4$ and $\ell\mu=6$ in~Figure\eqref{Entropy2}
and~Figure\eqref{Entropy1}, this result is nontrivial since the belt
width here is neither too large nor too small. The minimal surface
$\gamma_\mathcal{A}$ will not be very close to the horizon. Due to
the lake of numerical control at very low temperature, compared to
the critical temperature of the phase transition, we can not plot
all points in superconducting phase (the lowest temperature
in~Figure\eqref{Entropy2} and~Figure\eqref{Entropy1} is about
$0.64T_c$ and $0.68T_c$ separately). However, it is not obtrusive to
conclude that the entanglement entropy does behave as thermal
entropy at least for not very low temperature. Furthermore, as the
thermal entropy is equivalent to the black hole entropy in the
holographic setup, our calculation seems to support the viewpoint
that black hole entropy is due to the entanglement
entropy~\cite{Jacobson:1994iw,Kabat:1995eq,Solodukhin:2006xv,Emparan:2006ni}.


\section{Conclusion and discussions}
\label{sect:conclusion}

In a recent paper~\cite{Cai:2012sk}, we reported the behavior of
entanglement entropy in the holographical insulator/superconductor
phase transition, where a non-monotonic behavior of the entanglement
entropy was found  as the change of chemical potential. Because of
the absence of horizon in  the soliton background, both the
temperature and thermal entropy in dual boundary system do vanish.
It is unable to extract the relationship between entanglement
entropy and thermal entropy or black hole entropy. In this paper we
overcome the shortcoming by studying the holographic p-wave
superconductor at finite temperature. The other motivation is to see
how the behavior of the entanglement entropy will change as we tune
the parameter $\alpha$, the ratio of the gravitational constant to
the Yang-Mills coupling constant, which can change the order of the
phase transition.

In the fully back reacted case, we found $\frac{\rho_s}{\rho}$
scales as $T-T_c$ near critical point, which is the same as the
result in the probe limit~\cite{Gubser:2008wv}. As noted in
Ref.~\cite{Gubser:2008wv} this scaling behavior is different from
the transition between superconductivity and pseudogap state of high
$T_c$ materials, where the fraction is finite and nonvanishing. The
behavior is reminiscent of superfluid properties of
He~$\mathrm{II}$. However, the critical exponent in He~$\mathrm{II}$
is about $0.67$, while here it is one.

We found that the qualitative behavior of entanglement entropy is
dramatically different for sufficiently
 small and large $\alpha$ cases as we lower the temperature. For the case $\alpha<\alpha_c$,
 the entanglement entropy is continuous at the critical temperature $T_c$, while it has a jump for
  $\alpha>\alpha_c$ case. That is, the behavior of the entanglement entropy
  shows that the phase transition is second order when $\alpha
  <\alpha_c$, while it is first order as $\alpha >\alpha_c$.

  When the belt width is very large, i.e., $\gamma_\mathcal{A}$ probes deeply into the bulk,
  the entanglement entropy is extensive as the thermal entropy of the bulk black holes. In the opposite
   limit, the behavior
  perfectly  fits
  the general form obtained from four-dimensional conformal theories. Motivated by Ref.~\cite{Swingle:2011mk},
   there may exists a crossover function connecting the two limits. We compared the behaviors of the entanglement
   entropy and thermal entropy  during the whole process of the superconductor phase
transition.  It shows that they behave qualitatively the same.
 This result is nontrivial since the belt width here is neither too large nor too small
 (note that $\ell\mu=4$ and $\ell\mu=6$ in~Figure.\eqref{Entropy2}
and~Figure.\eqref{Entropy1}).  From the minimal surface picture, the
minimal surface $\gamma_\mathcal{A}$ can not be very close to
``hugging" the horizon at these intermediate scales. Due to the lack
of numerical control at low temperature, in our numerical
calculation we are not able to display the behavior of the
entanglement entropy  at sufficiently low temperature, compared to
the critical temperature of the phase transition.  As a result, we
can conclude that the entanglement entropy is a good probe
 to the holographic pase transition, and that its behavior can indicate not
 only the appearance, but also the order of the phase transition.

Here it is quite interesting to compare our results with those from
the case with holographic s-wave superconductors in
Ref.~\cite{Albash:2012pd}. The model studied there is a $SO(3)\times
SO(3)$ invariant truncation of four-dimensional ${\cal N}=8$ gauged
supergravity~\cite{Bobev:2011rv}. Different from other ``top-down"
models of holographic superconductors (see, for example,
Refs.~\cite{Gubser:2009qm,Gauntlett:2009dn,Aprile:2011uq}), where
the superconductor phase transition is a second order one, the phase
transition studied in Ref.~\cite{Albash:2012pd} will be second order
or first order, depending on the boundary condition of a scalar
field in the model.  In the second order phase transition case, the
entanglement entropy shows the feature of the second order phase
transition: the entanglement entropy is continuous and its slop has
a jump at the critical temperature.  On the order hand, in the case
of first order phase transition, there exists not only a jump for
the entanglement entropy at the critical temperature, but also a
kink shape (swallowtail shape) for the entanglement entropy for a
given range of strip widths in the superconducting phase. The kink
persists even for zero temperature solutions. It was argued there
that the appearance of the kink can be attributed to the existence
of a new scale in the theory and the entanglement entropy is a good
probe of the new scale. The new scale can be viewed as a finite
correlation length. Quite interestingly, in our p-wave
superconductor model, the kind does not appear even in the first
order phase transition case (see Figure.\ref{Entropywidth1}). This
suggests that the existence of this new scale is not universal in
 holographic superconductor models.
 Note that in this p-wave model, not only the $U(1)$ symmetry but also the rotational symmetry are broken, hence,
 the superconducting phase is anisotropic. On the other hand, only the $U(1)$ symmetry is broken in the s-wave model
 studied in Ref.~\cite{Albash:2012pd} and the superconducting phase is isotropic. We suspect that
  the anisotropy might be responsible for the absence of the kink in the superconducting phase in the
   first order transition case. Clearly it would be quite interesting to investigate whether such kink is
   common in other s-wave superconductor models.

Note that the zero temperature solution studied in
Ref.~\cite{Albash:2012pd} is an RG flow between two AdS spaces.
Indeed,  the existence of the kink for the entanglement entropy in
such backgrounds has also been observed in  Ref.~\cite{Myers:2012ed}
(see also Ref.~\cite{Liu:2012ee}) where
  null energy condition is used to constrain the monotonic behavior of c-function along RG flows. It was shown
  there that the kink would emerge for particular choice of the asymptotically AdS geometry, which is obviously
  shown in~Figure(6) in Ref.~\cite{Myers:2012ed}. More specifically, the geometry is controlled by  a parameter $R$.
   A kink will emerge for $R<R_c$, while it will disappear for the case $R>R_c$.
   In Ref.~\cite{Albash:2012pd}, it has been also analyzed in some
   details when the kink will appear. It shows that the presence of
   the kink is due to the particular potential in the model which in turn further determines the
    geometry structure of the background solutions. Let us further
  stress here that the kink of entanglement entropy also appears
    in the AdS soliton
    backgrounds~\cite{Nishioka:2006gr,Klebanov:2007ws,Cai:2012sk},
    and the signal of the kink acts as the ``confinement/deconfiment"
    phase transition.

Finally we would like to mention that the holographic p-wave
superconductor  model has been extended to include the Gauss-Bonnet
term in
 Refs.~\cite{Cai:2010cv,Cai:2010zm} where the condensate is found to become harder
  as the Gauss-Bonnet coefficient grows up. On the other hand, the original entanglement entropy
  proposal has been generalized to include some higher derivative corrections,
  such as the Gauss-Bonnet term, in the bulk~\cite{deBoer:2011wk,Hung:2011xb}. The entanglement entropy in
  the Gauss-Bonnet gravity has been
  calculated for many configurations and discovered to give additional
  contributions~\cite{Myers:2012ed,Ogawa:2011fw,Ishihara:2012jg}.
  There might be some new features for the entanglement entropy in the p-wave superconductor models with
  the Gauss-Bonnet term. We wish to report on the related work in  future.

\section*{Acknowledgements}

We would like to thank Hai-Qing Zhang, Zhang-Yu Nie, and Shu-Hao Zou for their helpful discussions and suggestions.
This work was supported in part by the National Natural Science Foundation of China (No.10821504, No.10975168
and No.11035008), and in part by the Ministry of Science and Technology of China under Grant No. 2010CB833004.

\appendix


\begin{thebibliography}{99}



\baselineskip 12pt
\bibitem{Maldacena:1997re}
  J.~M.~Maldacena,
  ``The large N limit of superconformal field theories and supergravity,''
  Adv.\ Theor.\ Math.\ Phys.\  {\bf 2}, 231 (1998)
  [Int.\ J.\ Theor.\ Phys.\  {\bf 38}, 1113 (1999)]
  [arXiv:hep-th/9711200].
\bibitem{Gubser:1998bc}
  S.~S.~Gubser, I.~R.~Klebanov and A.~M.~Polyakov,
  ``Gauge theory correlators from non-critical string theory,''
  Phys.\ Lett.\  B {\bf 428}, 105 (1998)
  [arXiv:hep-th/9802109].
\bibitem{Witten:1998qj}
  E.~Witten,
  ``Anti-de Sitter space and holography,''
  Adv.\ Theor.\ Math.\ Phys.\  {\bf 2}, 253 (1998)
  [arXiv:hep-th/9802150].


\bibitem{Herzog:2009xv}
  C.~P.~Herzog,
  ``Lectures on Holographic Superfluidity and Superconductivity,''  J.\ Phys.\ A A {\bf 42}, 343001 (2009)  [arXiv:0904.1975 [hep-th]].


\bibitem{Hartnoll:2007ih}
  S.~A.~Hartnoll, P.~K.~Kovtun, M.~Muller and S.~Sachdev,
  ``Theory of the Nernst effect near quantum phase transitions in condensed matter, and in dyonic black holes,''  Phys.\ Rev.\ B {\bf 76}, 144502 (2007)  [arXiv:0706.3215 [cond-mat.str-el]].

\bibitem{Liu:2009dm}
  H.~Liu, J.~McGreevy and D.~Vegh,
  ``Non-Fermi liquids from holography,''  Phys.\ Rev.\ D {\bf 83} (2011) 065029  [arXiv:0903.2477 [hep-th]].

\bibitem{Hartnoll:2009sz}
  S.~A.~Hartnoll,
  ``Lectures on holographic methods for condensed matter physics,''  Class.\ Quant.\ Grav.\  {\bf 26}, 224002 (2009)  [arXiv:0903.3246 [hep-th]].  

\bibitem{McGreevy:2009xe}
  J.~McGreevy,
  ``Holographic duality with a view toward many-body physics,''  Adv.\ High Energy Phys.\  {\bf 2010}, 723105 (2010)  [arXiv:0909.0518 [hep-th]].  

\bibitem{Sachdev:2012dq}
  S.~Sachdev,
  ``The Quantum phases of matter,''  arXiv:1203.4565 [hep-th].

\bibitem{Gubser:2008px}
  S.~S.~Gubser,
  ``Breaking an Abelian gauge symmetry near a black hole horizon,''
  Phys.\ Rev.\  D {\bf 78}, 065034 (2008)
  [arXiv:0801.2977 [hep-th]].

\bibitem{Hartnoll:2008vx}
  S.~A.~Hartnoll, C.~P.~Herzog and G.~T.~Horowitz,
  ``Building a Holographic Superconductor,''
  Phys.\ Rev.\ Lett.\  {\bf 101}, 031601 (2008)
  [arXiv:0803.3295 [hep-th]].

\bibitem{Nishioka:2009zj}
  T.~Nishioka, S.~Ryu and T.~Takayanagi,
  ``Holographic Superconductor/Insulator Transition at Zero Temperature,''  JHEP {\bf 1003}, 131 (2010)  [arXiv:0911.0962 [hep-th]].

\bibitem{Gubser:2008zu}
  S.~S.~Gubser,
  ``Colorful horizons with charge in anti-de Sitter space,''  Phys.\ Rev.\ Lett.\  {\bf 101}, 191601 (2008)  [arXiv:0803.3483 [hep-th]].

\bibitem{Gubser:2008wv}
  S.~S.~Gubser and S.~S.~Pufu,
  ``The Gravity dual of a p-wave superconductor,''  JHEP {\bf 0811}, 033 (2008)  [arXiv:0805.2960 [hep-th]].

\bibitem{2006PhRvB..73x5115R}
Ryu, S.,  Hatsugai, Y.
``Entanglement entropy and the Berry phase in the solid state,"  Phys.\ Rev.\  B {\bf 73}, 245115 (2006)
[arXiv:cond-mat/0601237].

\bibitem{Amico:2007ag}
  L.~Amico, R.~Fazio, A.~Osterloh and V.~Vedral,
  ``Entanglement in many-body systems,''  Rev.\ Mod.\ Phys.\  {\bf 80}, 517 (2008)  [quant-ph/0703044 [QUANT-PH]].


\bibitem{Ryu:2006bv}
  S.~Ryu and T.~Takayanagi,
  ``Holographic derivation of entanglement entropy from AdS/CFT,''  Phys.\ Rev.\ Lett.\  {\bf 96}, 181602 (2006)  [hep-th/0603001].

\bibitem{Nishioka:2009un}
  T.~Nishioka, S.~Ryu and T.~Takayanagi,
  ``Holographic Entanglement Entropy: An Overview,''  J.\ Phys.\ A A {\bf 42}, 504008 (2009)  [arXiv:0905.0932 [hep-th]].

\bibitem{Takayanagi:2012kg}
  T.~Takayanagi,
  ``Entanglement Entropy from a Holographic Viewpoint,''  arXiv:1204.2450 [gr-qc].


\bibitem{Albash:2008eh}
  T.~Albash and C.~V.~Johnson,
  ``A Holographic Superconductor in an External Magnetic Field,''  JHEP {\bf 0809}, 121 (2008)  [arXiv:0804.3466 [hep-th]].

\bibitem{Hartnoll:2008kx}
  S.~A.~Hartnoll, C.~P.~Herzog and G.~T.~Horowitz,
  ``Holographic Superconductors,''  JHEP {\bf 0812}, 015 (2008)  [arXiv:0810.1563 [hep-th]].

\bibitem{Cai:2011tm}
  R.~-G.~Cai, L.~Li, H.~-Q.~Zhang and Y.~-L.~Zhang,
  ``Magnetic Field Effect on the Phase Transition in AdS Soliton Spacetime,''  Phys.\ Rev.\ D {\bf 84}, 126008 (2011)  [arXiv:1109.5885 [hep-th]].  

\bibitem{Horowitz:2008bn}
  G.~T.~Horowitz and M.~M.~Roberts,
  ``Holographic Superconductors with Various Condensates,''  Phys.\ Rev.\ D {\bf 78}, 126008 (2008)  [arXiv:0810.1077 [hep-th]].

\bibitem{Brynjolfsson:2009ct}
  E.~J.~Brynjolfsson, U.~H.~Danielsson, L.~Thorlacius and T.~Zingg,
  ``Holographic Superconductors with Lifshitz Scaling,''  J.\ Phys.\ A A {\bf 43}, 065401 (2010)  [arXiv:0908.2611 [hep-th]].

\bibitem{Cai:2009hn}
  R.~-G.~Cai and H.~-Q.~Zhang,
  ``Holographic Superconductors with Horava-Lifshitz Black Holes,''  Phys.\ Rev.\ D {\bf 81}, 066003 (2010)  [arXiv:0911.4867 [hep-th]].

\bibitem{Horowitz:2009ij}
  G.~T.~Horowitz and M.~M.~Roberts,
  ``Zero Temperature Limit of Holographic Superconductors,''  JHEP {\bf 0911}, 015 (2009)  [arXiv:0908.3677 [hep-th]].

\bibitem{Horowitz:2011dz}
  G.~T.~Horowitz, J.~E.~Santos and B.~Way,
  ``A Holographic Josephson Junction,''  Phys.\ Rev.\ Lett.\  {\bf 106}, 221601 (2011)  [arXiv:1101.3326 [hep-th]].


\bibitem{Montull:2011im}
  M.~Montull, O.~Pujolas, A.~Salvio and P.~J.~Silva,
  ``Flux Periodicities and Quantum Hair on Holographic Superconductors,''  Phys.\ Rev.\ Lett.\  {\bf 107}, 181601 (2011)  [arXiv:1105.5392 [hep-th]].  


\bibitem{Bobev:2011rv}
  N.~Bobev, A.~Kundu, K.~Pilch and N.~P.~Warner,
  ``Minimal Holographic Superconductors from Maximal Supergravity,''  JHEP {\bf 1203}, 064 (2012)  [arXiv:1110.3454 [hep-th]].


\bibitem{Liu:2012hc}
  Y.~Liu, Y.~Peng and B.~Wang,
  ``Gauss-Bonnet holographic superconductors in Born-Infeld electrodynamics with backreactions,''  arXiv:1202.3586 [hep-th].

\bibitem{Erdmenger:2011tj}
  J.~Erdmenger, P.~Kerner and H.~Zeller,
  ``Transport in Anisotropic Superfluids: A Holographic Description,''  JHEP {\bf 1201}, 059 (2012)  [arXiv:1110.0007 [hep-th]].


\bibitem{Albash:2012pd}
  T.~Albash and C.~V.~Johnson,
  ``Holographic Studies of Entanglement Entropy in Superconductors,''  arXiv:1202.2605 [hep-th].

\bibitem{Cai:2012sk}
  R.~-G.~Cai, S.~He, L.~Li and Y.~-L.~Zhang,
  ``Holographic Entanglement Entropy in Insulator/Superconductor Transition,''  arXiv:1203.6620 [hep-th].


\bibitem{Ammon:2009xh}
  M.~Ammon, J.~Erdmenger, V.~Grass, P.~Kerner and A.~O'Bannon,
  ``On Holographic p-wave Superfluids with Back-reaction,''  Phys.\ Lett.\ B {\bf 686}, 192 (2010)  [arXiv:0912.3515 [hep-th]].



\bibitem{Witten:1998zw}
  E.~Witten,
  ``Anti-de Sitter space, thermal phase transition, and confinement in gauge theories,''  Adv.\ Theor.\ Math.\ Phys.\  {\bf 2}, 505 (1998)  [hep-th/9803131].


\bibitem{Srednicki:1993im}
  M.~Srednicki,
  ``Entropy and area,''  Phys.\ Rev.\ Lett.\  {\bf 71}, 666 (1993)  [hep-th/9303048].

\bibitem{Casini:2005zv}
  H.~Casini and M.~Huerta,
  ``Entanglement and alpha entropies for a massive scalar field in two dimensions,''  J.\ Stat.\ Mech.\  {\bf 0512}, P12012 (2005)  [cond-mat/0511014].  

\bibitem{Nishioka:2006gr}
  T.~Nishioka and T.~Takayanagi,
  ``AdS Bubbles, Entropy and Closed String Tachyons,''  JHEP {\bf 0701}, 090 (2007)  [hep-th/0611035].











\bibitem{Swingle:2011mk}
  B.~Swingle and T.~Senthil,
  ``Universal crossovers between entanglement entropy and thermal entropy,''  arXiv:1112.1069 [cond-mat.str-el].

\bibitem{Jacobson:1994iw}
  T.~Jacobson,
  ``Black hole entropy and induced gravity,''  gr-qc/9404039.

\bibitem{Kabat:1995eq}
  D.~N.~Kabat,
  ``Black hole entropy and entropy of entanglement,''  Nucl.\ Phys.\ B {\bf 453}, 281 (1995)  [hep-th/9503016].


\bibitem{Solodukhin:2006xv}
  S.~N.~Solodukhin,
  ``Entanglement entropy of black holes and AdS/CFT correspondence,''  Phys.\ Rev.\ Lett.\  {\bf 97}, 201601 (2006)  [hep-th/0606205].

\bibitem{Emparan:2006ni}
  R.~Emparan,
  ``Black hole entropy as entanglement entropy: A Holographic derivation,''  JHEP {\bf 0606}, 012 (2006)  [hep-th/0603081].

\bibitem{Bobev:2011rv}
  N.~Bobev, A.~Kundu, K.~Pilch and N.~P.~Warner,
  ``Minimal Holographic Superconductors from Maximal Supergravity,''  JHEP {\bf 1203}, 064 (2012)  [arXiv:1110.3454 [hep-th]].

\bibitem{Gubser:2009qm}
  S.~S.~Gubser, C.~P.~Herzog, S.~S.~Pufu and T.~Tesileanu,
  ``Superconductors from Superstrings,''  Phys.\ Rev.\ Lett.\  {\bf 103}, 141601 (2009)  [arXiv:0907.3510 [hep-th]].

\bibitem{Gauntlett:2009dn}
  J.~P.~Gauntlett, J.~Sonner and T.~Wiseman,
  ``Holographic superconductivity in M-Theory,''  Phys.\ Rev.\ Lett.\  {\bf 103}, 151601 (2009)  [arXiv:0907.3796 [hep-th]].


\bibitem{Aprile:2011uq}
  F.~Aprile, D.~Roest and J.~G.~Russo,
  ``Holographic Superconductors from Gauged Supergravity,''  JHEP {\bf 1106}, 040 (2011)  [arXiv:1104.4473 [hep-th]].

\bibitem{Myers:2012ed}
  R.~C.~Myers and A.~Singh,
  ``Comments on Holographic Entanglement Entropy and RG Flows,''  JHEP {\bf 1204}, 122 (2012)  [arXiv:1202.2068 [hep-th]].

\bibitem{Liu:2012ee}
  H.~Liu and M.~Mezei,
``A Refinement of entanglement entropy and the number of degrees of
freedom,''
  arXiv:1202.2070 [hep-th].


\bibitem{Klebanov:2007ws}
  I.~R.~Klebanov, D.~Kutasov and A.~Murugan,
``Entanglement as a probe of confinement,''
  Nucl.\ Phys.\ B {\bf 796}, 274 (2008)
  [arXiv:0709.2140 [hep-th]].

\bibitem{Cai:2010cv}
  R.~-G.~Cai, Z.~-Y.~Nie and H.~-Q.~Zhang,
  ``Holographic p-wave superconductors from Gauss-Bonnet gravity,''  Phys.\ Rev.\ D {\bf 82}, 066007 (2010)  [arXiv:1007.3321 [hep-th]].

\bibitem{Cai:2010zm}
  R.~-G.~Cai, Z.~-Y.~Nie and H.~-Q.~Zhang,
  ``Holographic Phase Transitions of P-wave Superconductors in Gauss-Bonnet Gravity with Back-reaction,''  Phys.\ Rev.\ D {\bf 83}, 066013 (2011)  [arXiv:1012.5559 [hep-th]].  


\bibitem{deBoer:2011wk}
  J.~de Boer, M.~Kulaxizi and A.~Parnachev,
  ``Holographic Entanglement Entropy in Lovelock Gravities,''  JHEP {\bf 1107}, 109 (2011)  [arXiv:1101.5781 [hep-th]].

\bibitem{Hung:2011xb}
  L.~-Y.~Hung, R.~C.~Myers and M.~Smolkin,
  ``On Holographic Entanglement Entropy and Higher Curvature Gravity,''  JHEP {\bf 1104}, 025 (2011)  [arXiv:1101.5813 [hep-th]].

\bibitem{Ogawa:2011fw}
  N.~Ogawa and T.~Takayanagi,
  ``Higher Derivative Corrections to Holographic Entanglement Entropy for AdS Solitons,''  JHEP {\bf 1110}, 147 (2011)  [arXiv:1107.4363 [hep-th]].  

\bibitem{Ishihara:2012jg}
  M.~Ishihara, F.~-L.~Lin and B.~Ning,
  ``Refined Holographic Entanglement Entropy for the AdS Solitons and AdS black Holes,''  arXiv:1203.6153 [hep-th].



\end{thebibliography}
\end{document}